\documentstyle[12pt,epsfig,rotate]{article}

\newcommand{\gsim}{\;\rlap{\lower 3.5 pt \hbox{$\mathchar \sim$}} \raise
1pt \hbox {$>$}\;}
\newcommand{\lsim}{\;\rlap{\lower 3.5 pt \hbox{$\mathchar \sim$}} \raise
1pt \hbox {$<$}\;}

\def\kpnn{K^+\rightarrow\pi^+\nu\bar\nu}

\def\klpn{K_{\rm L}\rightarrow\pi^0\nu\bar\nu}

\newcommand{\bea}{\begin{eqnarray}}
\newcommand{\eea}{\end{eqnarray}}
\newcommand{\bd}{\begin{displaymath}}
\newcommand{\ed}{\end{displaymath}}
\newcommand{\be}{\begin{equation}}
\newcommand{\ee}{\end{equation}}
\newcommand{\ord}{{\cal O}}

\newcommand{\epe}{\varepsilon'/\varepsilon}
\newcommand{\mt}{m_{\rm t}}

\newcommand{\mw}{M_{\rm W}}

\newcommand{\vcb}{|V_{cb}|}
\newcommand{\vtd}{|V_{td}|}
\newcommand{\vts}{|V_{ts}|}
\newcommand{\vub}{|V_{ub}/V_{cb}|}
\newcommand{\vus}{|V_{us}|}

\newcommand{\beq}{\begin{equation}}
\newcommand{\eeq}{\end{equation}}

\begin{document}



\author{ {\large\bf A.J.~Buras${}^{1}$, P. Gambino${}^{2}$,
    M. Gorbahn${}^{1}$,}  \\
  {\large\bf S. J\"ager${}^{1}$
    and L. Silvestrini${}^{1,3}$} \\
  \ \\
  {\small\bf ${}^{1}$ Physik Department,
    Technische Universit\"at M\"unchen,} \\
  {\small\bf D-85748 Garching, Germany} \\
  {\small\bf ${}^{2}$ Theory Division, CERN, CH-1211 Geneva 23, 
    Switzerland} \\
  {\small\bf ${}^{3}$ Dipartimento di Fisica, Universit\`a di Roma
    ``La Sapienza'' and} \\
  {\small\bf INFN, Sezione di Roma, P.le A. Moro, I-00185, Roma, Italy}
}

\date{}
\title{
{\normalsize\sf
\rightline{TUM-HEP-379/00}
\rightline{CERN-TH/2000-190}
}
\bigskip
{\LARGE\bf
Universal Unitarity Triangle and \\
Physics Beyond the Standard Model 
}}

\maketitle
\thispagestyle{empty}

\phantom{xxx} \vspace{-9mm}

\begin{abstract}
  We make the simple observation that there exists a {\it universal
    unitarity triangle} for all models, like the SM, the Two Higgs
  Doublet Models I and II and the MSSM with minimal flavour violation,
  that do not have any new operators beyond those present in the SM
  and in which all flavour changing transitions are governed by the
  CKM matrix with no new phases beyond the CKM phase. This universal
  triangle can be determined in the near future from the ratio
  $(\Delta M)_d/(\Delta M)_s$ and $\sin 2\beta$ measured first through
  the CP asymmetry in $B_d^0\to \psi K_S$ and later in
  $K\to\pi\nu\bar\nu$ decays.  Also suitable ratios of the branching
  ratios for $B\to X_{d,s}\nu\bar\nu$ and $B_{d,s}\to\mu^+\mu^-$ and
  the angle $\gamma$ measured by means of CP asymmetries in B decays
  can be used for this determination.  Comparison of this universal
  triangle with the non-universal triangles extracted in each model
  using $\varepsilon$, $(\Delta M)_d$ and various branching ratios for
  rare decays will allow to find out in a transparent manner which of
  these models, if any, is singled out by experiment. A virtue of the
  universal triangle is that it allows to separate the determination
  of the CKM parameters from the determination of new parameters
  present in the extensions of the SM considered here.
 \end{abstract}

\newpage
\setcounter{page}{1}
\setcounter{footnote}{0}

\section{Introduction}
\setcounter{equation}{0}
One of the important goals of particle physics is the determination of
the Cabibbo--Kobayashi--Maskawa (CKM) matrix \cite{CAB,KM}.
In addition to the leading tree level 
K and B decays, flavour changing neutral current processes generated 
at the one loop
level in the Standard Model (SM) and sensitive to the top quark couplings
$V_{t d(s)}$ play a crucial role in this determination. This program
is not only complicated by the presence of hadronic uncertainties but
also by the possible existence of new physics that contributes to
various quantities through diagrams involving new particles.
These new contributions depend on unknown parameters, like the masses
and couplings of new particles, that pollute the extraction of the CKM
parameters.

We would like to point out that in a certain class of extensions of 
the SM it is possible to
construct measurable quantities that depend on the CKM parameters
but are not polluted by new physics contributions. This means that
these quantities allow a direct determination of the ``true'' values
of the CKM parameters which are common to the SM and this particular class
of its extensions. Correspondingly there exists a universal unitarity
triangle common to all these models. Interestingly the quantities
required to construct the universal unitarity triangle are essentially
free from hadronic uncertainties.

In order to explain our point we use the Wolfenstein parameterization 
\cite{WO} of the CKM matrix and its generalization to include
higher order terms in $\lambda$ \cite{BLO}.

Let us recall first that
the four Wolfenstein parameters $\lambda$, $A$, $\varrho$ and $\eta$
can be determined in the standard manner as follows:

{\bf Step 1:}

The parameters $\lambda$ and $A$ are determined from semileptonic K and B
decays sensitive to the elements $\vus$ and $\vcb$ respectively:
\be
\lambda=\vus=0.22, \qquad  A=\frac{\vcb}{\lambda^2}=0.826\pm0.041~.
\ee
As the decays in question are tree level decays with large
branching ratios this determination is to an excellent approximation
independent of any possible physics beyond the SM.

{\bf Step 2:}

The parameters $\varrho$ and $\eta$ are determined by constructing
with the help of various decays the unitarity triangle of fig.~1, 
where \cite{BLO}
\begin{equation}\label{CKM4}
\bar\varrho=\varrho (1-\frac{\lambda^2}{2})~,
\qquad
\bar\eta=\eta (1-\frac{\lambda^2}{2})
\end{equation}
describe the apex of this triangle.
The lengths CB, CA and BA are equal respectively to
\begin{equation}\label{2.94a}
1\,, \qquad R_b \equiv  \sqrt{\bar\varrho^2 +\bar\eta^2}
= (1-\frac{\lambda^2}{2})\frac{1}{\lambda}
\left| \frac{V_{ub}}{V_{cb}} \right|\,,
\qquad
R_t \equiv \sqrt{(1-\bar\varrho)^2 +\bar\eta^2}
=\frac{1}{\lambda} \left| \frac{V_{td}}{V_{cb}} \right|.
\end{equation}

\begin{figure}   
    \begin{center}
\input{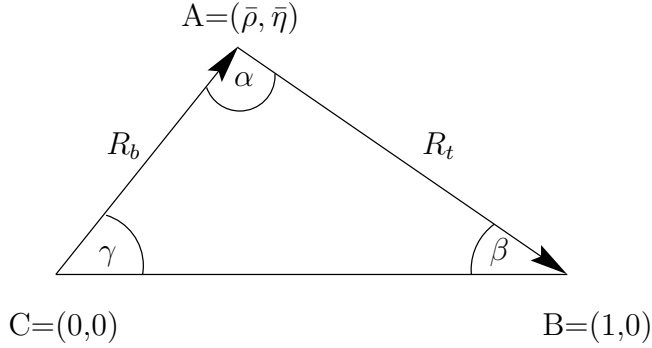}
    \end{center}
    \caption[]{Unitarity Triangle.}
    \label{fig:utriangle}
\end{figure}

The standard construction of this triangle involves the ratio
$\vub$ extracted from inclusive and exclusive tree level B decays
and flavour changing neutral current processes such as $B^0_d-\bar B^0_d$
mixing (the mass difference $(\Delta M)_d$) and indirect CP violation 
in $K_L$ decays (the parameter $\varepsilon$), both sensitive to the
CKM element $V_{td}$. There is also a constraint coming from the
lower bound on the mass difference $(\Delta M)_s$ describing 
$B^0_s-\bar B^0_s$ mixing. In particular in the case of 
$B^0_{d,s}-\bar B^0_{d,s}$ mixings the following formulae for
$(\Delta M)_{d,s}$ resulting from box diagrams are used:

\begin{equation}
(\Delta M)_{d,s} = \frac{G_F^2}{6 \pi^2} \eta_B m_{B_{d,s}} 
(\hat B_{B_{d,s}} F_{B_{d,s}}^2 ) M_W^2 F_{tt} |V_{t(d,s)}|^2
\label{eq:xds}
\end{equation}
Here $F_{tt}$ is a function of $\mt$ and $\mw$
 resulting from box diagrams with top
quark exchanges,
$\hat B_B$ is a non-perturbative parameter, $F_B$ is
the B meson decay constant and 
$\eta_B$ the short distance QCD factor \cite{NLOS1,NLOS3} common
to $(\Delta M)_{d}$ and $(\Delta M)_{s}$.

Similarly, the experimental value for $\varepsilon$ combined
with the theoretical calculation of box diagrams describing 
$K^0-\bar K^0$ mixing gives the constraint for
$(\bar\varrho,\bar\eta)$ in the form of the following 
hyperbola \cite{AJBLH}:
\begin{equation}\label{100}
\bar\eta \left[(1-\bar\varrho) A^2 \eta_2 F_{tt}
+ P_c(\varepsilon) \right] A^2 \hat B_K = 0.226~.
\end{equation}
Here $\hat B_K$ is a  non-perturbative parameter analogous to
$\hat B_{B_{d,s}}$, $\eta_2$ is a short distance
QCD correction \cite{NLOS1}, $F_{tt}$ is the function
present also in (\ref{eq:xds}) and
$P_c(\varepsilon) =0.31\pm0.05$ \cite{NLOS2}  summarizes 
charm--charm and charm--top contributions.

Combining the two steps above one can determine the
range of values of $(\bar\varrho,\bar\eta)$ consistent with all present
data. Analyses of this type can be found in
\cite{AJBLH,ALI00,Ciuchini:2000xh,Parodi}. In the future this
procedure can be generalized to include CP asymmetries in B decays
sensitive to the angles of the unitarity triangle and various
branching ratios for K and B decays sensitive to the sides and the
height of this triangle \cite{AJBLH}.  If the SM is the correct theory
all these measurements should result in a unique value of
$(\bar\varrho,\bar\eta)$.

This procedure of testing the SM can be applied to its extensions
as well.
Step 1 remains unchanged as this determination, based on
tree level decays, is insensitive to physics beyond the
SM.  
On the other hand Step 2 can be affected by new physics
due to:
\begin{itemize}
\item
New contributions to box diagrams modifying the function
$F_{tt}$ and to the analogous functions describing various penguin
diagrams contributing to rare K and B decays.
This introduces new parameters into the box function $F_{tt}$ and
the penguin functions that in the SM depend only on $\mt$ and $\mw$.
\item 
New contributions to box and penguin diagrams that are not
  proportional to the same combination of CKM matrix elements as the
  SM top contribution (for example, new contributions to $P_c$ in
  eq.~(\ref{100}) or new contributions to $(\Delta M)_{d,s}$
  proportional to $|V_{cd(s)}^*V_{cb}|^2$).
\item
New complex phases beyond the one present in the CKM matrix.
\item
New local operators contributing to the relevant amplitudes
beyond those present in the SM. This would introduce additional
non-perturbative factors $B_i$ and new box and penguin functions.
\end{itemize} 
It is evident from (\ref{eq:xds}) and (\ref{100}) that any
modification of the function $F_{tt}$ will change the values of the
extracted $(\bar\varrho,\bar\eta)$. A recent analysis of this type
within the MSSM can be found in \cite{ALI00}.  Similar comments apply
to the extraction of $(\bar\varrho,\bar\eta)$ from various branching
ratios for rare K and B decays.  Moreover if new phases are present in
the extensions of the SM, CP violating asymmetries will generally
measure different quantities than $\alpha$, $\beta$ and
$\gamma$ in fig.~1.  For instance the CP asymmetry in 
$B\to \psi K_S$ will no
longer measure $\beta$ but $\beta+\theta_{NP}$ where $\theta_{NP}$ is
a new phase. Strategies for dealing with such situations have been
developed. See for instance \cite{NIR96,BNEW} and references therein.

The presence of new physics and of new phases will be signaled by
inconsistencies in the $(\bar\varrho,\bar\eta)$ plane.  In order to
sort out which type of new physics is responsible for deviations from
the SM expectations one has to study many loop induced decays and many
CP asymmetries. Some ideas in this direction can be found in
\cite{NIR96,BNEW}.

While in principle a global fit of all experimental data
can be used to test the SM and its
extensions it is desirable to develop strategies which allow to make
these tests in a transparent manner.

Here we will concentrate on models like the SM, the Two Higgs Doublet
Models (TDHM) I and II and the MSSM with minimal flavour violation,
that do not have any new operators beyond those present in the SM
\cite{TAN} and in which all flavour changing transitions are governed
by the CKM matrix with no new phases beyond the CKM phase.
Furthermore, in these models the only sizable new contributions are
proportional to the same CKM parameters as the SM top contributions.
That is, only the values of the functions describing top-mediated
contributions to box and penguin diagrams are modified.

We would like to point out that the models in this class
share a useful  property. Namely, the CKM parameters in these models 
extracted  from a particular set of data are independent of the contributing
loop functions like $F_{tt}$, they are universal in this class of models. 
Correspondingly there exists a {\it universal unitarity triangle}.
The determination of this universal unitarity triangle and of the 
corresponding CKM
parameters has four virtues:

\begin{itemize}
\item
The CKM matrix can be determined without the knowledge of new
unknown parameters present in these particular extensions of the SM.
\item
Because the extracted CKM matrix is also valid in these 
models, the dependence of various quantities on the new parameters 
becomes more transparent. In short: the determination of the CKM 
matrix and of the new parameters 
can be separated from each other, as opposed to the present strategies
discussed in step 2 above.
\item
The comparison of the predictions for a given observable in the SM
and in this kind of extensions can then be done keeping the CKM
parameters fixed.
\item
The extraction of the universal CKM parameters
is essentially free from hadronic uncertainties.
\end{itemize}

In what follows we will list the set of quantities which allow a
determination of the universal unitarity triangle. Subsequently we
will indicate how the models in this class can be distinguished
from each other and from more complicated models which
bring in new complex phases and new operators.
\section{Determination of $R_t$}
In order to illustrate our point let us consider (\ref{eq:xds}).
Using this formula  one finds
\begin{equation}\label{107x}
\frac{\vtd}{|V_{ts}|}= 
\xi\sqrt{\frac{m_{B_s}}{m_{B_d}}}
\sqrt{\frac{(\Delta M)_d}{(\Delta M)_s}}\equiv\kappa,
\qquad
\xi = 
\frac{F_{B_s} \sqrt{\hat B_{B_s}}}{F_{B_d} \sqrt{\hat B_{B_d}}}.
\end{equation}
This ratio depends only on measurable quantities 
$(\Delta M)_{d,s}$, $m_{B_{d,s}}$ and the non-perturbative parameter
$\xi$. Now to an excellent accuracy \cite{BLO}
\be\label{vts1}
\vtd=\vcb\lambda R_t, \qquad 
\vts=\vcb(1-\frac{1}{2}\lambda^2+\bar\varrho\lambda^2)
\ee
with $\bar\varrho$ defined in (\ref{CKM4}).
We note next that through the unitarity of the CKM
matrix, the present experimental upper bound on 
$(\Delta M)_d/(\Delta M)_s$ and the value of $\vub$
 one has in all these models $0\le\bar\varrho\lsim 0.5$, where
$\xi=1.16\pm 0.07$ \cite{Becirevic:2000nv} has been used. 
Consequently
$\vts$ deviates from $\vcb$ by at most $2.5\%$. This means
that to a very good accuracy $R_t$ is given by
\be\label{Rt}
R_t=\frac{\kappa}{\lambda}
\ee
independently of new parameters characteristic for a given model 
and of $\mt$.
If necessary the $\ord(\lambda^2)$ corrections in (\ref{vts1}) can
be incorporated in (\ref{Rt}). This will be only required when the
error on $\xi$ will be decreased below $2\%$, which is clearly
a very difficult task.

While the ratio $(\Delta M)_d/(\Delta M)_s$ will be the first one to serve
our purposes, there are at least two other quantities which allow
a clean measurement of $R_t$ within the class of extensions of the
SM considered.
These are the ratios
\begin{equation}\label{bxnn}
\frac{Br(B\to X_d\nu\bar\nu)}{Br(B\to X_s\nu\bar\nu)}=
\left|\frac{V_{td}}{V_{ts}}\right|^2
\end{equation}
\begin{equation}\label{bmumu}
\frac{Br(B_d\to\mu^+\mu^-)}{Br(B_s\to\mu^+\mu^-)}=
\frac{\tau_{B_d}}{\tau_{B_s}}\frac{m_{B_d}}{m_{B_s}}
\frac{F^2_{B_d}}{F^2_{B_s}}
\left|\frac{V_{td}}{V_{ts}}\right|^2
\end{equation}
which similarly to $(\Delta M)_d/(\Delta M)_s$ measure 
\begin{equation}\label{vtdts}
\left|\frac{V_{td}}{V_{ts}}\right|^2=\lambda^2
\frac{(1-\bar\varrho)^2+\bar\eta^2}{1+\lambda^2(2\bar\varrho-1)}
\approx \lambda^2 R_t^2~.
\end{equation}
Out of these three ratios the cleanest is (\ref{bxnn}), which is
essentially free of hadronic uncertainties \cite{GBGI}. Next comes
(\ref{bmumu}), involving $SU(3)$ breaking effects in the ratio of $B$
meson decay constants.  Finally, $SU(3)$ breaking in the ratio of bag
parameters $\hat B_{B_d}/\hat B_{B_s}$ enters in addition in
(\ref{107x}). These $SU(3)$ breaking effects should eventually be
calculable with reasonable precision from lattice QCD.

It should be remarked that the branching ratio for the rare decay 
$\kpnn$ is known to provide a clean measurement of $V_{td}$ and
consequently of $R_t$ \cite{AJBLH}. However, this branching ratio alone
cannot serve our purposes because it is sensitive to new physics 
contributions.

\section{Determination of $\beta$ and $\gamma$}
In order to complete the determination of $\bar\varrho$ and $\bar\eta$
in the universal unitarity triangle 
one can use $\sin2\beta$ extracted either from the CP asymmetry
in $B_d\to\psi K_S$ \cite{SANDA} or from $K\to\pi\nu\bar\nu$ decays 
\cite{BB4}.
In the first case one has to measure the time dependent asymmetry
\begin{equation}\label{best}
a_{CP}(t,\psi K_S)= -\sin(2\beta) \sin((\Delta M)_dt) 
\end{equation}
that allows a measurement of the angle $\beta$ 
without any hadronic uncertainties.
In the second case the measurements of $Br(\kpnn)$ and $Br(\klpn)$ 
are required.
Then \cite{BB4}:
\begin{equation}\label{imre}
\sin 2\beta=\frac{2 r_s}{1+r^2_s}
\end{equation}
with
\begin{equation}\label{cbb}
r_s(B_1, B_2)=\sqrt{\sigma}{\sqrt{\sigma(B_1-B_2)}-
P_c(\nu\bar\nu)\over\sqrt{B_2}}\,.
\end{equation}
Here $\sigma=1/(1-\lambda^2/2)^2$ and $B_{1,2}$ stand for
 the ``reduced'' branching ratios
\begin{equation}\label{b1b2}
B_1={Br(\kpnn)\over 4.11\cdot 10^{-11}}\qquad
B_2={Br(\klpn)\over 1.80\cdot 10^{-10}}\,.
\end{equation}
It should be stressed that $\sin 2\beta$ determined in this manner depends
only on two measurable branching ratios and on 
$P_c(\nu\bar\nu)=0.42\pm 0.06$ which is completely calculable 
in perturbation theory
\cite{BB2}.
Moreover, hadronic uncertainties in these decays have been found
to be negligibly small \cite{RS,GBGI}. As analyzed in \cite{BB4},
a measurement of both branching ratios within $\pm 10\%$ will
allow the determination of $\sin 2\beta$ within $\pm 0.05$.

Both extractions of $\sin 2\beta$ are to an excellent accuracy
independent of the new parameters characteristic for a given model.
In particular $P_c(\nu\bar\nu)$ being proportional to $V^*_{cs}
V_{cd}$ receives only negligible new contributions in the class of
models considered \cite{TAN}.

Concerning the determination of the angle $\gamma$,
the two theoretically cleanest methods 
are: i) the full time dependent analysis of 
$B_s\to D^+_s K^{-}$ and $\bar B_s\to D^-_s K^{+}$  \cite{adk}
and ii) the well known triangle construction due to Gronau and Wyler 
\cite{Wyler}
which uses six decay rates $B^{\pm}\to D^0_{CP} K^{\pm}$,
$B^+ \to D^0 K^+,~ \bar D^0 K^+$ and  $B^- \to D^0 K^-,~ \bar D^0 K^-$.
Variants of the latter method
which could be more promising experimentally
 have been proposed in \cite{DUN2,V97}.
Both methods involve only tree diagrams and 
are  unaffected by new physics contributions in
the class of models considered. 
It appears that these methods will give useful results at later stages
of CP-B investigations. In particular the first method will be feasible
only at LHC-B. Clearly any other method for the determination of $\gamma$
in which new physics of the type considered here can be eliminated
could also be used. For a recent review of $\gamma$ determinations
we refer to \cite{BF97} and references therein. 
\section{Determination of the Universal Unitarity Triangle}
Once $R_t$ and $\sin 2\beta$
have been determined as discussed above, $\bar\varrho$ and $\bar\eta$
can be found through \cite{B95}
\begin{equation}\label{5u}
\bar\eta=a\frac{R_t}{\sqrt{2}}\sqrt{\sin 2\beta \cdot r_{-b}(\sin 2\beta)}\,,
\quad\quad
\bar\varrho = 1-\bar\eta r_{b}(\sin 2\beta)\,
\end{equation}
where 
\be
r_b(z)=(1+b\sqrt{1-z^2})/z, \qquad a,b=\pm~.
\ee
Thus for given values of $(R_t,\sin 2\beta)$ there are four solutions
for  $(\bar\varrho,\bar\eta)$ corresponding to $(a,b)=(+,+),~(+,-),
~(-,+),~(-,-)$.
As described in \cite{B95}
three of these solutions can be eliminated by using further information, 
for instance coming from 
$|V_{ub}/V_{cb}|$ and $\varepsilon$, so that eventually the solution 
corresponding to $(a,b)=(+,+)$ is singled out
\begin{equation}\label{5a}
\bar\eta=\frac{R_t}{\sqrt{2}}\sqrt{\sin 2\beta \cdot r_{-}(\sin 2\beta)}\,,
\quad\quad
\bar\varrho = 1-\bar\eta r_{+}(\sin 2\beta)\,~.
\end{equation}
We will illustrate this with an example below.

On the other hand $\bar\varrho$ and $\bar\eta$ following from
$R_t$ and $\gamma$ are simply given by
\begin{equation}\label{6a}
\bar\eta=R_b \sin\gamma
\quad\quad
\bar\varrho = R_b \cos\gamma
\end{equation}
with
\begin{equation}\label{7a}
R_b=\cos\gamma\pm\sqrt{R^2_t-\sin^2\gamma}.
\end{equation}
Comparing the resulting $R_b$ with the one extracted from $\vub$
 (see (\ref{2.94a})) one of the two solutions can be eliminated.

As an alternative to $\sin 2\beta$ or $\gamma$ one could use the measurement
of $\sqrt{\bar\varrho^2+\bar\eta^2}$ by means of $\vub$ but this
strategy suffers from hadronic uncertainties in the extraction
of $\vub$. Similarly using $\vub$ and $\gamma$ one can construct the
the universal unitarity triangle by means of (\ref{6a}).

We observe that all these different methods
determine the ``true'' values of 
$\bar\eta$ and $\bar\varrho$ independently of 
new physics contributions in
the class of models considered.
Since $\lambda$ and $\vcb=A\lambda^2$ are determined from tree level
K and B decays they are insensitive to new physics as well. Thus
the full CKM matrix can be determined in this manner. The corresponding
universal unitarity triangle common to all the models considered
can be found directly from  formulae like (\ref{5u}), (\ref{5a})
and (\ref{6a}). 

\begin{table}
\begin{center}
\begin{tabular}{|c||c|c|c|c|}\hline
$(a,b)$& $\bar\eta$ &$\bar\varrho$&$\gamma$& $R_b$\\ \hline
$(+,+)$ & $0.35$ &$ 0.15$ &$ 67^\circ$& $0.38$\\ \hline
$(+,-)$ & $0.85$ &$ 0.65$ &$ 53^\circ$& $1.07$\\ \hline
$(-,+)$ & $-0.35$ &$ 1.85$ &$ -11^\circ$ &$1.88$\\ \hline
$(-,-)$ & $-0.85$ &$ 1.35$ &$ -32^\circ$ &$1.59$\\ \hline
\end{tabular}
\end{center}
\centerline{} 
\caption{Four solutions for $\bar\eta$ and $\bar\varrho$ using $R_t$ 
and $\sin(2\beta)$.} 
\label{TAB6}
\end{table}

As an example let us take 
$(\Delta M)_d=0.471/ps$, $(\Delta M)_s=16.0/ps$ and $\xi=1.16$.
This gives $R_t=0.92$. Taking $\sin 2\beta=0.70$ one finds then by means
of (\ref{5u}) the four solutions for the universal unitarity
triangle given in table~1.
As from the data on $\vub$ we have $R_b\lsim 0.5$ only 
the first solution is allowed.

Concentrating on the allowed solution, in table 2 we illustrate with a
few examples the accuracy of the determination of the unitarity
triangle.  The first two rows give the assumed input parameters and
their experimental errors. The remaining rows give the results for
$\bar\eta$, $\bar\varrho$, $\gamma$ and $R_b$ where errors have been
added in quadrature.

The accuracy in the scenario I 
 should be achieved at B-factories, FNAL and HERA-B. 
 Scenarios II and
III correspond to B-physics at Fermilab during 
the Main Injector era, LHC-B and BTeV.
It should be stressed that this high accuracy is achieved not only  
because of our assumptions about future experimental errors in the
scenarios considered, but also because 
of the clean character of the quantities considered.

\begin{table}
\begin{center}
\begin{tabular}{|c||c|c|c|c|}\hline
& Central &$I$&$II$&$III$\\ \hline
$R_t$ & $0.92$ &$\pm 0.10$ &$\pm 0.05$ & $\pm 0.03 $\\ \hline
$\sin(2\beta)$ & $0.70$ &$\pm 0.06$ &$\pm 0.02$ & $\pm 0.01 $\\ \hline
\hline
$\bar\eta$ &$0.348$ &$\pm 0.052$ &$\pm 0.022$&$\pm 0.013$ \\ \hline
$\bar\varrho$ &$0.148$ &$\pm 0.094$&$\pm 0.047$ &$\pm 0.028$\\ \hline
$\gamma$ &$66.9^\circ$ &$\pm 15.2^\circ$&$\pm 7.6^\circ$ &
$\pm 4.6^\circ$\\ \hline
$R_b$ &$0.378$ &$\pm 0.039$ &$\pm 0.013$&$\pm 0.006$
 \\ \hline
\end{tabular}
\end{center}
\centerline{} 
\caption{Determinations of $\bar\eta$ and $\bar\varrho$ using $R_t$ 
and $\sin(2\beta)$.} 
\label{TAB5}
\end{table}

Having the allowed values of table~1 at hand one can calculate $\varepsilon$,
$\epe$, $(\Delta M)_d$, $(\Delta M)_s$ and branching ratios for
rare decays. As these quantities depend on the parameters characteristic
for a given model the results for the SM, the MSSM and other models
of this class will generally differ from
each other. Consequently by comparing these predictions with the
data one will be able to find out which
of these  models is singled out by experiment.
Equivalently, $\varepsilon$,
$\epe$, $(\Delta M)_d$, $(\Delta M)_s$ and branching ratios for
rare decays allow to determine non-universal unitarity triangles
that depend on the model considered. Only those unitarity
triangles which are the same as the universal triangle survive
the test. 

It is of course possible that  new physics is more complicated than
discussed here and that new complex phases and new operators
beyond those present in the SM have to be taken into account.
These types of effects would be signaled by:
\begin{itemize}
\item
Inconsistencies between different constructions of the universal
triangle,
\item
Disagreements of the data with the $(\Delta M)_{d,s}$ and the branching
ratios for rare K and B decays predicted on the basis of the
universal unitarity triangle for all models of the class considered
here.
\end{itemize}

In our opinion the universal unitarity triangle provides 
a transparent strategy to distinguish between models belonging to
the class considered in this paper and to  search for physics
beyond the SM. Its other virtues have been listed at the end of
the Introduction. 
Presently we do not know this triangle as all the
available measurements used for the construction of the unitarity
triangle are sensitive to  physics beyond the SM. 
It is exciting, however, that in the coming years this triangle will be
known once $(\Delta M)_s$ has been measured and $\sin 2\beta$ extracted 
from the CP asymmetry in $B_d^0\to \psi K_S$. At later stages 
$K\to\pi\nu\bar\nu$,
$B\to X_{d,s}\nu\bar\nu$, $B_{d,s}\to \mu^+\mu^-$
 and future determinations of $\gamma$ through CP asymmetries in
B decays will also be very useful in this respect.

This work has been supported in part by the German Bundesministerium 
f\"ur Bildung and Forschung under the contract 05HT9WOA0.

\vfill\eject


\begin{thebibliography}{99}
\bibitem{TAN}
We assume that $\tan\beta$ is not too large so that additional
operators that become relevant for very large $\tan\beta$ can
be neglected. It should also be noted that in the MSSM with minimal
flavour violation, all contributions that are not proportional to
$V_{td(s)}$ are suppressed by $m_u$ or $m_c$ and therefore they are
numerically negligible. The same is true for the charm-$H^+$
contributions in the THDM I and II.
\end{thebibliography}
\end{document}